# GHz-bandwidth InAs/InAsSbP barrier infrared detectors for the 3.0–3.7 μm spectral region operating at room temperature

**Grzegorz Gomółka,[1, a)] Krzysztof Kłos,[2] Jarosław Pawluczyk,[3] Maciej Kowalczyk,[1] Piotr Martyniuk,[3] and Łukasz A. Sterczewski[1, b)]**

## AFFILIATIONS

[1] Laser & Fiber Electronics Group, Faculty of Electronics, Photonics and Microsystems, Wroclaw University of Science and Technology, Wybrzeze Wyspianskiego 27, 50-370 Wroclaw, Poland
[2] Photin sp. z o.o, 15 Lutosławskiego St., 05-080 Klaudyn, Poland
[3] Institute of Applied Physics, Military University of Technology, 2 Kaliskiego St., 00-908 Warsaw, Poland

[a)] **Author to whom correspondence should be addressed:** grzegorz.gomolka@pwr.edu.pl
[b)] **Electronic mail:** lukasz.sterczewski@pwr.edu.pl

## ABSTRACT

The demand for fast mid-wave infrared photodetectors is fueled by high-rate free-space optical communication and optical frequency comb spectroscopy. To date, only a few multi-GHz photodetectors have shown sensitive room-temperature operation in the 3.0-3.7 μm band, yet their commercial availability remains scarce. In this work, we present the remarkable response speed of an InAs/InAsSbP *pBn* barrier detector – a type typically not associated with high-frequency operation. A weakly reverse-biased photodiode with a diameter of 121 μm achieves a −3 dB electrical bandwidth of 2.4 GHz and −20 dB bandwidth of 8.0 GHz. This is the best result in this class of mid-infrared photodetectors confirmed optically. High signal-to-noise photodetection is also demonstrated at frequencies exceeding 19 GHz. The relatively simple device structure (devoid of cascaded structure or type-II superlattice) was realized on the mature InAs material platform, which opens new perspectives for accessible, sensitive, multi-GHz photodetectors for the 3.0–3.7 μm spectral region.

In recent years, mid-wave infrared (MWIR, 3-5 μm wavelength) photodetectors have experienced remarkable growth driven by numerous applications, including chemical sensing, imaging, and defense. Some fields like free-space optical telecommunications (FSO)[1,2] or frequency comb spectroscopy and metrology[3,4] require high-speed photodetection in the MWIR spectral region with electrical bandwidths (BW) exceeding 1 GHz. Unfortunately, achieving such performance is still challenging for room-temperature devices, with successful demonstrations limited to leading academic and industrial facilities.[5] The most widespread and mature technology of HgCdTe (MCT) photodetectors still struggles to reach a single-GHz BW in ambient conditions. Among high-speed room-temperature photoreceivers for this band, quantum cascade detectors (QCDs) hold the speed record.[6] Their unipolar electron transport and picosecond-scale intersubband relaxation times yield tens-of-GHz BWs in the long-wavelength part of the MWIR and beyond. In 2021, QCD technology reached commercial maturity with 20 GHz devices offered by Hammamatsu.[7,8] While QCDs are perfect for multi-Gbit/s FSO links in the atmospheric window near 4.1 μm, their responsivity reaches only single mA/W, and they cease to operate at wavelengths below 3.4 μm, relevant for chemical analysis and spectroscopy. In this band, interband cascade infrared photodetectors (ICIPs) with few-GHz BWs have been reported, providing coverage from the short-wave infrared (SWIR) to ~5 μm.[9–12] Despite their slower carrier dynamics than QCDs, ICIPs have also been successfully applied in FSO at 4.1 μm,[13] as well as in precise dual-comb spectroscopy.[14,15] Their vital advantage over QCDs is also a two orders of magnitude higher responsivity. Another competitive candidate for GHz-speed MWIR receivers are uni-travelling-carrier (UTC) photodetectors.[16–18] They have been reported with a similar spectral coverage, BW, and responsivity to ICIPs, with the notable demonstration of a type-II superlattice (T2SL) based UTC reaching a −3 dB BW of 12.8 GHz.[19] Several other MWIR sensing devices have also been presented with GHz BWs across the years, such as *p-i-n* InSb-based photodiode,[20] quantum well infrared photodetector,[21] microwave-resonator-assisted InSb detector,[22] or narrowband yet highly-sensitive resonant cavity infrared detector.[23,24] Despite these advances, a significant technological gap remains for multi-GHz, broadband, sensitive MWIR detectors with no commercially available solutions between 2.6 and 3.4 μm.

A group of infrared photodetectors typically not associated with high-speed response are barrier photodetectors.[25] In the MWIR, their dominant material system is currently a InAs/InAsSb T2SL, known for its stable uniform epitaxy, compatibility with InAs, GaAs and GaSb substrates, flexible band structure engineering, low Auger recombination, and excellent room-temperature performance.[26–30] Barrier devices prioritize dark current suppression and increased detectivity, which results in relatively thick absorber and barrier layers, as well as large sensor size. Moreover, they are often designed to operate at zero bias without optimization for radio frequency (RF) characteristics. Their physics does not, however, fundamentally limit the response speed.[31] Only very recently, InAsSb barrier photodetectors have been characterized electrically in terms of their frequency response using a vector network analyzer,[32,] and the first discussion of their application in FSO communication has been raised.[33]

In this work, we experimentally demonstrate the GHz operation of a fast InAs/InAsSbP *pBn* barrier photodetector in the 3.0–3.7 μm spectral range. The designed device resembles *p*-*i*-*n* with an extra barrier where the carriers are swept out by a high electric field, thus providing high frequency response. A Ø121 μm mesa illuminated through the substrate was wire bonded to a microwave transmission line and reverse biased to detect optical signals at frequencies exceeding 19 GHz. The measured electrical properties of the device agree well with the equivalent circuit model.

Infrared detectors reported here were grown on a 2" *n*-doped InAs wafer in a horizontal MOVPE reactor in a process similar to that described in Ref.[34] The structure, following the *pBn*-type barrier photodetector design, is shown schematically in Fig. 1(a), while the band diagram under reverse bias is plotted in Fig. 1(b). The unintentionally *n*-doped (UID) InAs MWIR absorption region (absorber, layer #2 in Fig. 1) is sandwiched between the InAsSbP wide-gap contact for holes, which is an electron-blocking barrier (layer #3), and the InAs heavily *n*-doped bottom contact for electrons (layer #1). In such a design, the placement of the *p*—*n* junction within a wide-bandgap barrier enables an expansion of the space-charge region across the absorber, unlike in typical barrier detectors.[35] This facilitates rapid removal of photocarriers by drift under reverse bias. A graded Zn *p*-type doping profile was implemented within the barrier layer, starting from a low concentration at the absorber interface. The lattice-matched quaternary InAsSbP material makes it possible to use a relatively thick barrier layer of 210 nm. The heavily doped *p*-type InAs contact layer (#4) above the barrier enables a near-ohmic metal contact at the top layer. The structure was also passivated with a 50-nm-thick $SiO_2$ layer. After growth and processing, the wafer was diced into sample plates comprising 11 mesas with diameters ranging from 121 to 1137 μm arranged on the 2.5 × 2.5 × 0.5 mm InAs substrate. In this study, we put emphasis on achieving high-speed MWIR signal detection, so the smallest Ø121 μm InAs mesa was selected for subsequent characterization (due to the lowest junction capacitance).

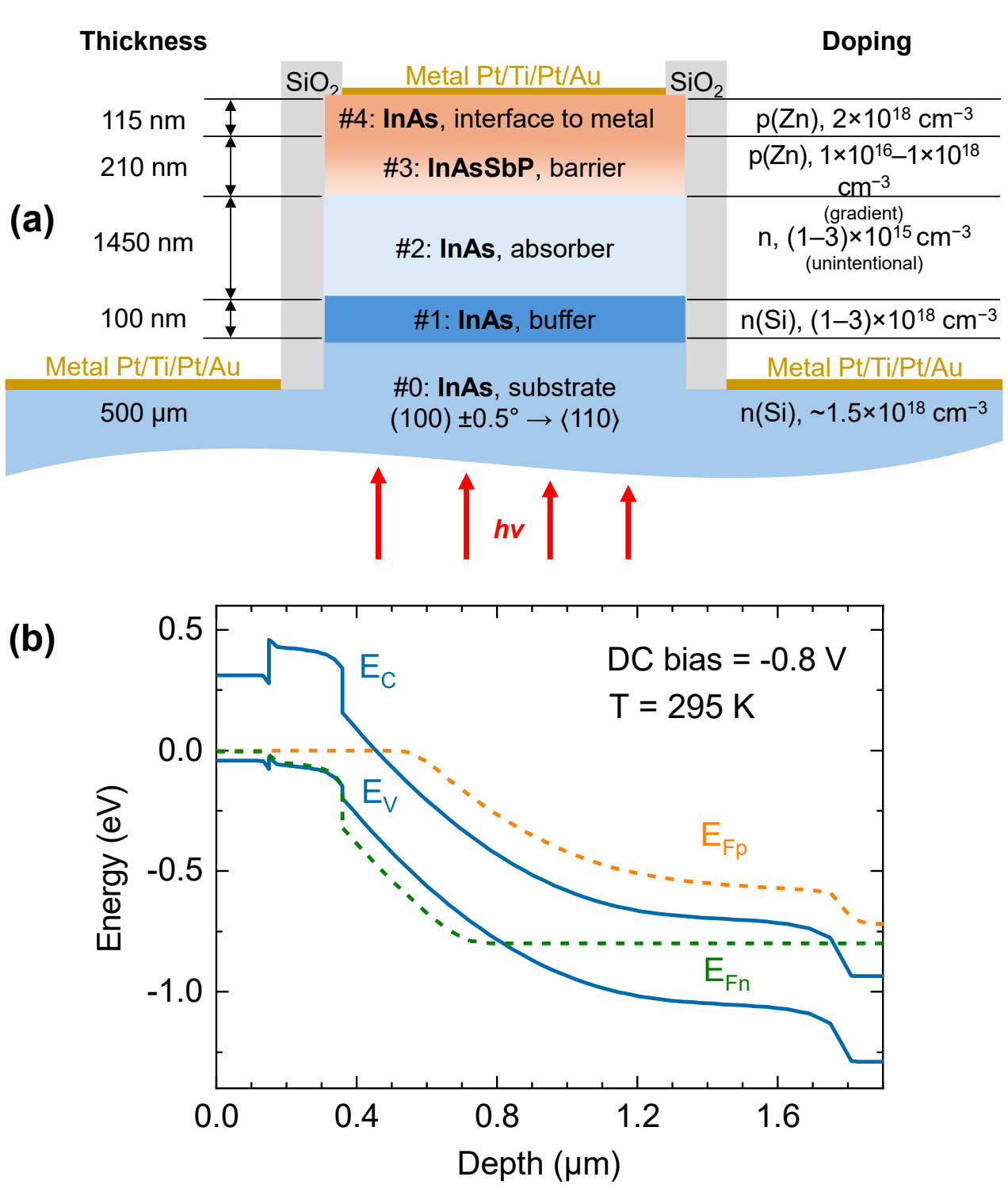

**FIG. 1.** (a) Schematic of the *pBn*-type InAs photodetector structure with layer dimensions and doping concentrations. (b) Band diagram of the detector under -0.8 V reverse bias. $E_{C/V}$ – conduction/valence band, $E_{Fn/p}$ – Fermi level for electrons/holes. Depth is measured from top contact down.

The quantum efficiency (QE) and responsivity curves of the analyzed device are shown in Fig. 2. They were measured using a Fourier transform infrared spectrometer (FTIR) with a thermal source. Under front illumination from the epi-layer side (dashed lines in Fig. 2), the QE of the InAs structure reached about 40% at the peak of its spectral response (0.8 A/W) located near 2.5 µm at 295 K. However, when fast MWIR detection is concerned, front light coupling is often not practical due to mesa miniaturization and microwave-optimization of the metal contacts, which either block or partially obscure light access from the top. Transparent conductive materials for MWIR are another challenge, as the transparency of regular indium tin oxide (ITO) is limited to ~2.5 µm, and novel materials are still under development.[36] Hence, back-illumination is typically employed in fast infrared detectors, as most of the semiconductor substrates are good MWIR transmitters, albeit inducing ~30% Fresnel loss due to the high substrate refractive index.

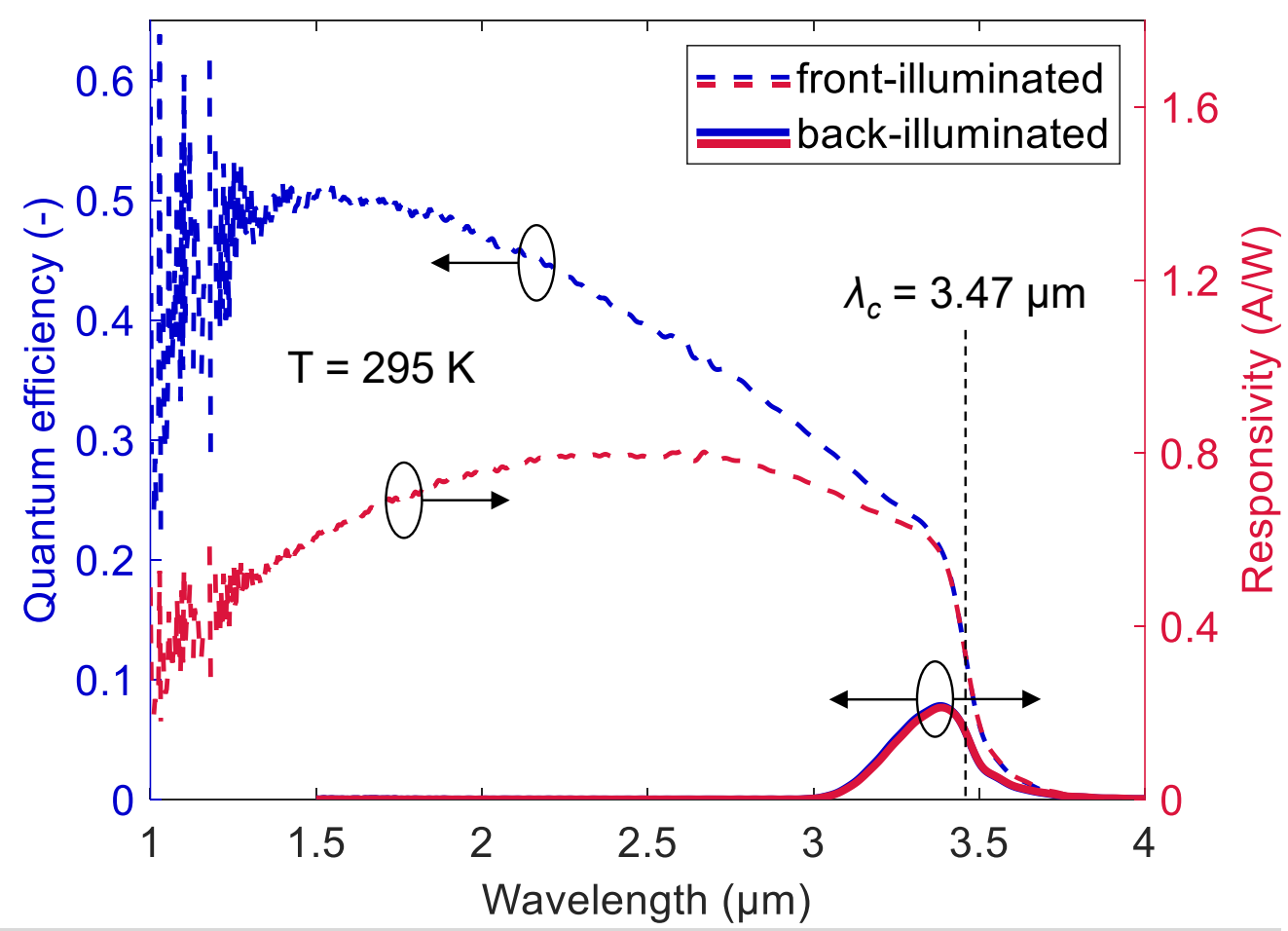

**FIG. 2.** Spectral QE and responsivity of the InAs *pBn* detector at 295 K: for front-illumination from the epi-layer side (dashed lines), and for back-illumination through the InAs substrate (solid lines). Note: QE and responsivity curves for back-illuminated structure overlap almost perfectly, hence double arrow axes indicator was used.

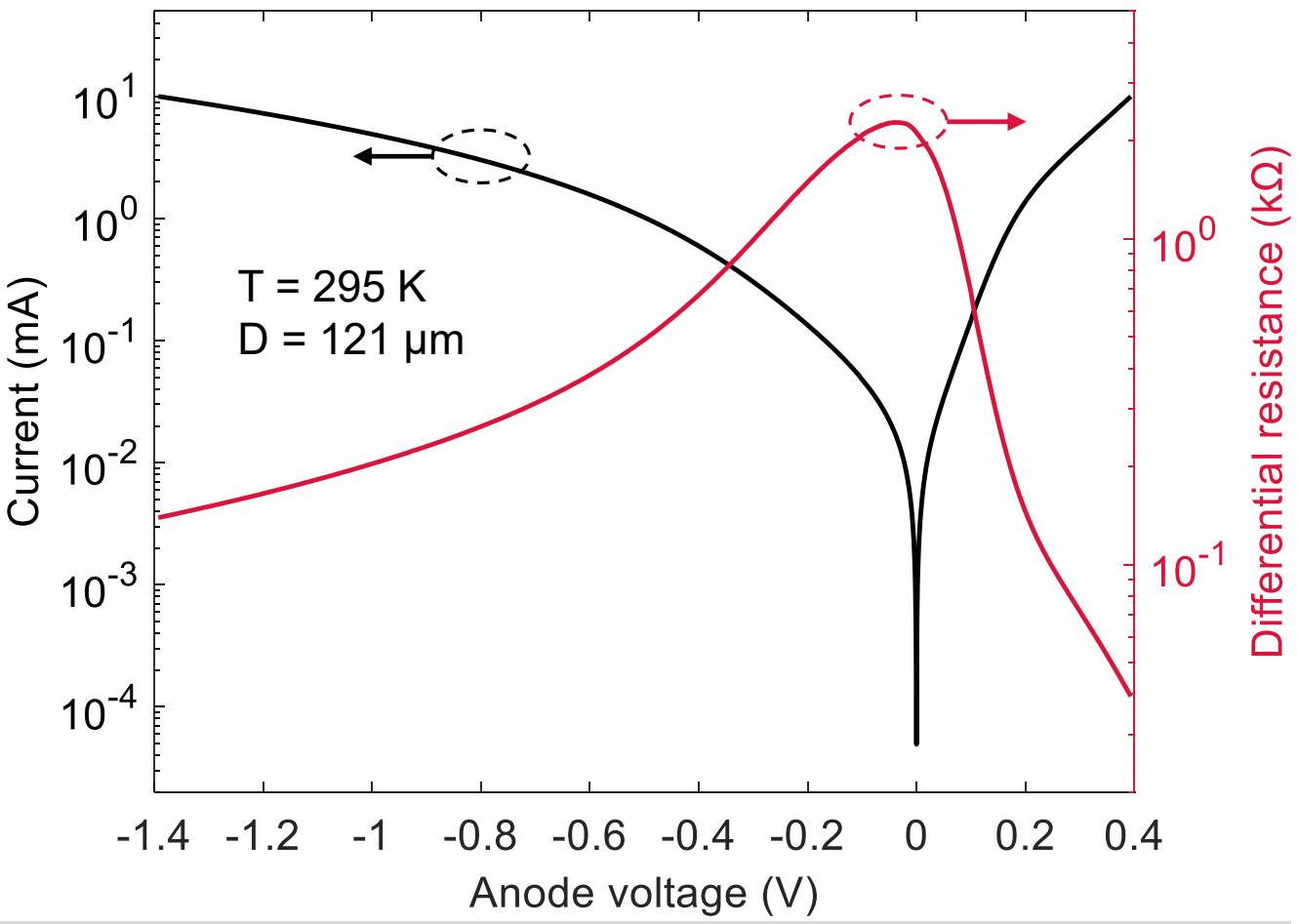

**FIG. 3.** I-V curve showing the absolute dark current of the Ø121 µm InAs photodiode as a function of bias voltage measured at 295 K. The differential resistance vs. bias was also plotted and referenced to the secondary y-axis.

When the InAs photodetector was illuminated from the substrate side (solid lines in Fig. 2), both band-to-band absorption and Urbach-tail behavior of the absorption coefficient were observed near the cut-off wavelength $\lambda_c$ of 3.47 µm. The deviation from the ideal photon-counter spectral shape results from the relatively small thickness of the absorber layer. Illumination through the highly $n$-doped InAs substrate is affected by the Burstein-Moss effect, which reduces the absorption coefficient near the bandgap and shifts the cut-off wavelength toward shorter values. This effect allows a small fraction of longer-wavelength light to penetrate the highly doped bottom layers and reach the low-doped absorber. Consequently, the overall QE and responsivity of the device are substantially exacerbated by absorption within the substrate, as shown in Fig. 2. Nevertheless, even when illuminated from the substrate side, InAs photodiode yields ~8% QE and 0.21 A/W responsivity at 3.4 µm at 295 K.

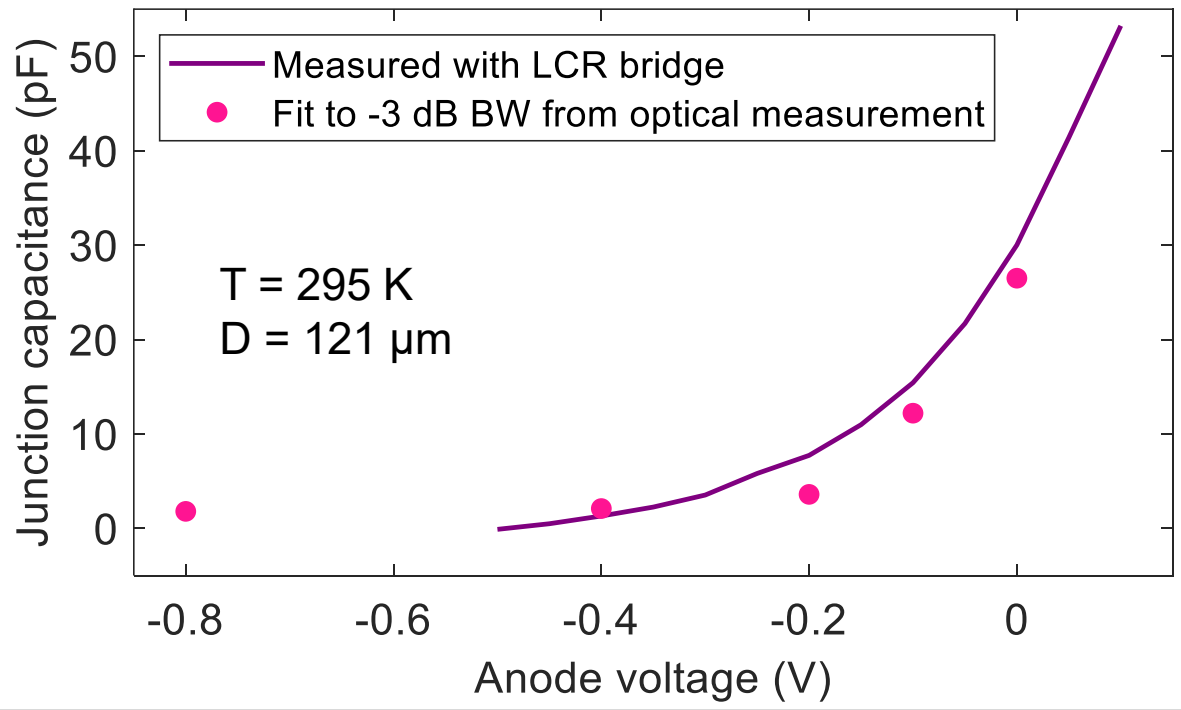

**FIG. 4.** Junction capacitance of the Ø121 µm InAs photodiode vs. bias voltage measured at 295 K. Solid line represents capacitance values measured with an LCR bridge corrected for the parasitic capacitance of the probe leads (ended at bias of -0.5 V due to reaching the accuracy limit of the device) and the dots are capacitance values obtained from the equivalent circuit model, set to match –3 dB BW values from optical measurements (see Fig. 6 and 7).

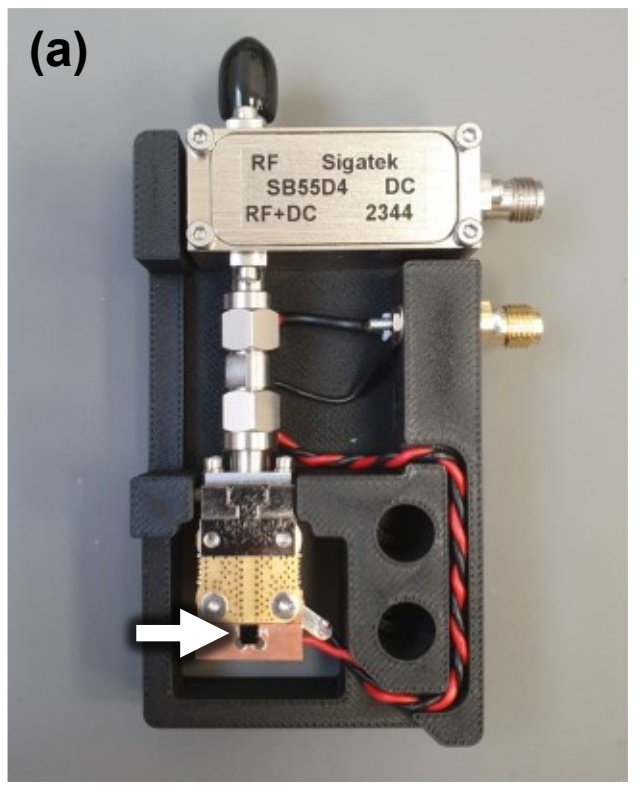

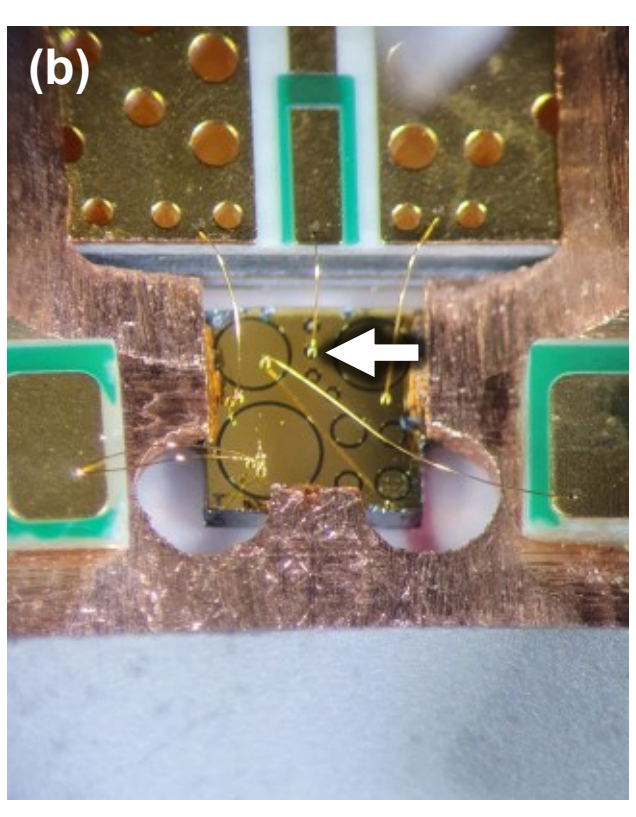

**FIG. 5.** (a) Photograph of the mounted InAs photodetector (the white arrow shows the semiconductor sample seen from the substrate side); (b) micrograph of the top contacts of the InAs photodiodes (the white arrow points the Ø121 µm mesa bonded to the GCPW seen above). Note: mesas bonded to the side pads were not used in this work.

The $I$-$V$ curve of the selected unilluminated photodiode was measured at room temperature using a source measure unit and is shown in Fig. 3. The bias voltage was limited to not exceed ±10 mA of dark current. Tunneling (trap-assisted) current dominates in the reverse bias range, beginning from zero volts. Based on the responsivity value of 0.21 A/W obtained at 3.4 µm and the Johnson noise at zero bias, the low-frequency specific detectivity $D^*$ of $1.2 \times 10^9$ cm·$\sqrt{\mathrm{Hz}}$/W was achieved for the unbiased, back-illuminated photodetector.

The capacitance of the photodiode junction was characterized using a programmable LCR bridge. The leads of the InAs detector were connected to the Kelvin test probes of the meter, which applied DC bias and a small 200 kHz AC signal to the structure to measure its complex impedance, from which the junction capacitance was derived. The measurement was then corrected for the parasitic capacitance of the probe leads, which was estimated to be ~10.3 pF. Data superimposed with fitted capacitance values for the optically measured −3 dB BW are presented in Fig. 4. Satisfactory agreement was achieved down to -0.5 V of DC bias. Below this value, the LCR bridge provided inaccurate results.

Assembly of the microwave-optimized fast InAs photodetector involved wire bonding of one Ø121 µm InAs photodiode to a grounded coplanar waveguide with a 25 µm gold wire in a ground-signal-ground (GSG) configuration.

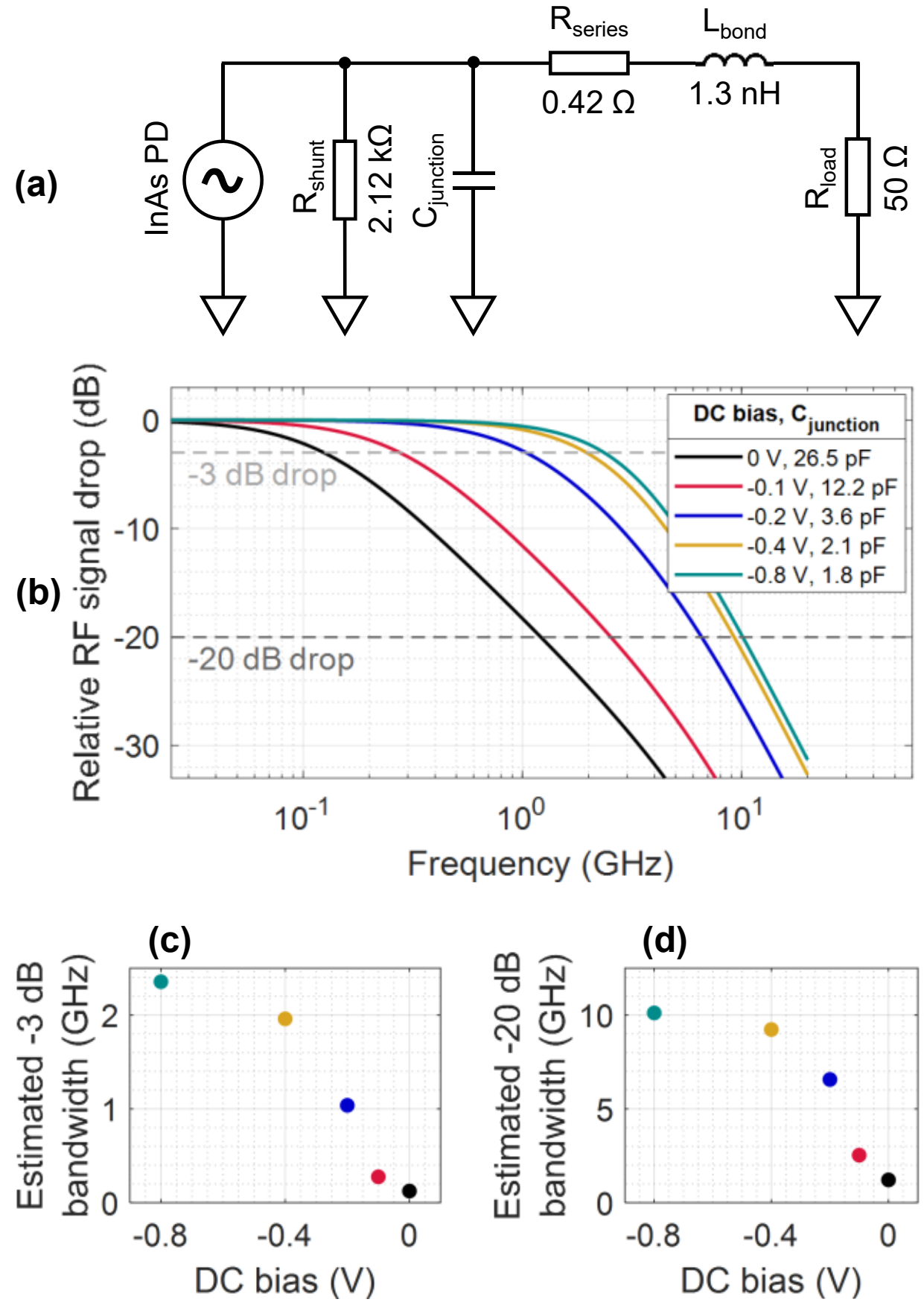

**FIG. 6.** (a) Equivalent circuit of the InAs photodetector. (b) Simulated frequency response for the circuit composed of an InAs photodetector and 50 Ω load. Note: junction capacitance value was changed in each simulation to reflect the biasing effects and set to fit the -3 dB BW obtained in optical response measurements (see Fig. 7(b)). (c) Simulated -3 dB and (d) -20 dB BW of InAs photodetector.

The other end of the waveguide was terminated with a 2.92 mm end launch RF connector. The component was subsequently connected to a GHz-speed bias tee to enable reverse biasing of the photodiode and microwave signal extraction. Simultaneously, the two larger mesas from the sample were bonded to additional side pads – they were not used in this study. The detector circuit mounted in a 3D-printed scaffold is shown in Fig. 5(a) with a zoom-in on the bonding arrangement in Fig. 5(b).

The InAs photodetector can be approximated by an equivalent circuit model consisting of an AC source, shunt and series resistors, a capacitor representing the junction capacitance, and an inductor representing the wire bond inductance (circuit shown in Fig. 6(a)). This model allows us to simulate the device's microwave frequency response. Experimentally, the shunt resistance was estimated from the I-V curve (linear fit within ±0.01 V of bias voltage), while the series resistance was obtained by fitting the I-V data at a larger forward bias (over 0.3 V) to the Shockley diode model. The inductance of the gold wire was estimated based on its length (1 nH/mm). The circuit was terminated with a 50 Ω test load for a frequency sweep. The junction capacitance value was varied to match the −3 dB BW obtained in the optical response measurement (see Fig. 7(b)). Response curves obtained for several reverse bias voltages are shown in Fig. 6(b), and the −3 dB and −20 dB BW values are shown in Fig. 6(c) and (d), respectively.

To experimentally assess the true electrical BW of the InAs photodetector under MWIR excitation, it was illuminated with a mid-infrared laser system based on a spectrally-broadened output of a Kerr-lens mode-locked Cr:ZnS laser covering wavelengths up to 4 µm.[37,38] Since its emission peak is located at wavelengths shorter than the responsivity range of the detector, a long pass filter with a cut-off wavelength of 2.4 µm was used, followed by a gray filter to reduce its average optical power to ~2 mW. The beam was subsequently focused with a black diamond aspheric lens on the Ø121 µm InAs mesa. The photodiode was reverse-biased by a programmable power supply through a 40 GHz bias tee whose RF port was connected to a radio frequency spectrum analyzer (RFSA) using a 1-m-long 2.92 mm semi-rigid coaxial cable. An internal preamplifier of the RFSA was employed to boost the RF signal from the photodetector.

The mid-infrared laser emitted ~25 fs pulse train with a 25 MHz repetition rate, which corresponded to a comb-like pattern in the frequency domain. The RF spectrum (impulse response) recorded for the highest tested photodiode bias of -0.8 V is presented in Fig. 7(a). The drop of the RF signal magnitude at multiples of the repetition frequency relative to the highest comb peak is shown in Fig. 7(b). The detector operating in photovoltaic mode is already perfectly capable of receiving optical signals at tens-to-hundreds of MHz with an estimated -3 dB BW of ~125 MHz. Biasing of the structure with even a small reverse voltage significantly increases the response speed. With just -0.2 V, the BW is extended to 1.1 GHz. The best performance was achieved for −0.8 V, with a −3 dB BW of 2.4 GHz and −20 dB BW of 8.0 GHz, respectively. Both 3 dB and 20 dB BWs tend to saturate versus voltage, indicating that the electric field extends over a region comparable to the absorber thickness. Further biasing showed little improvement in response speed, and above -1 V of bias, a substantial rise of the dark current was noted.

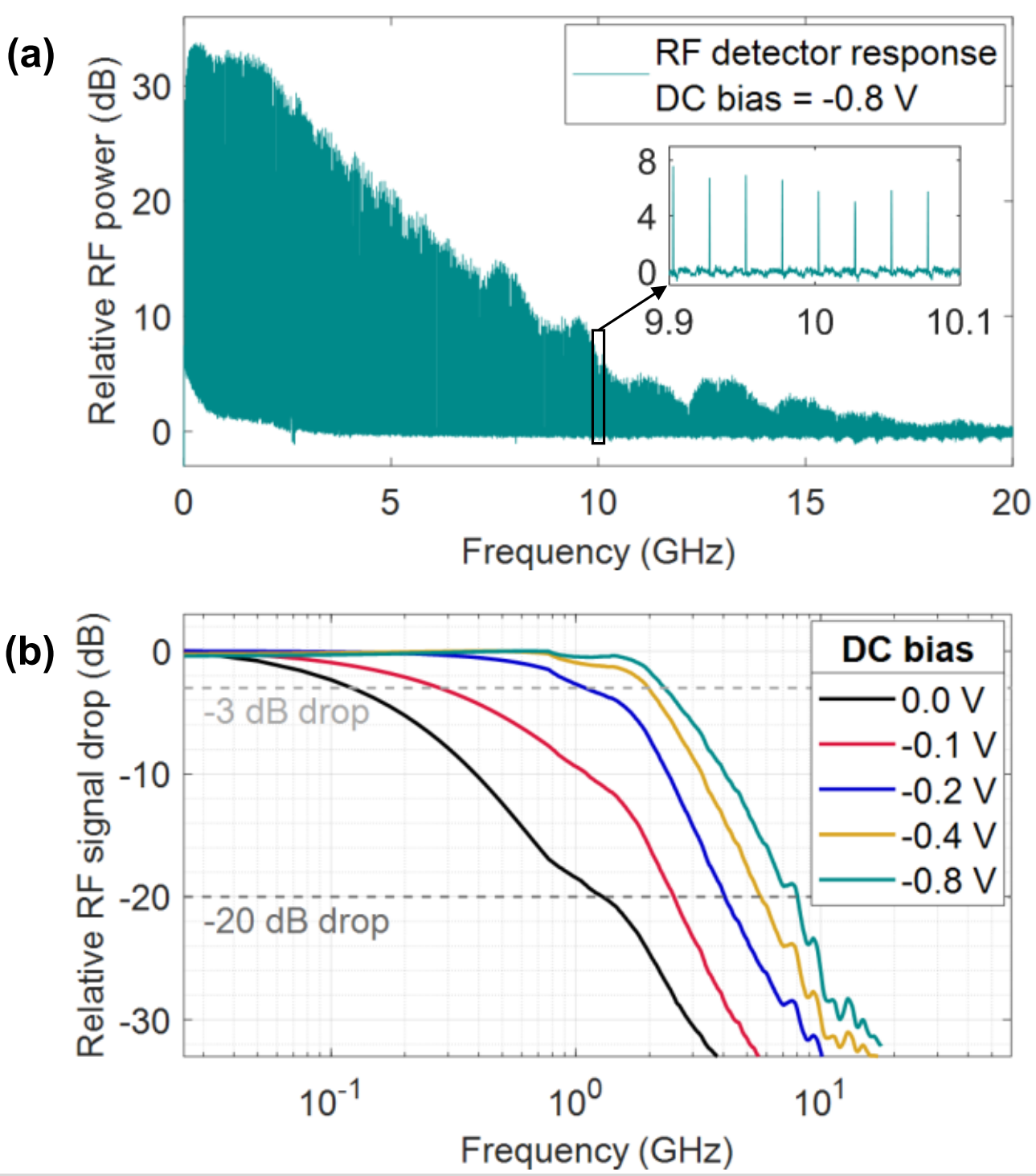

**FIG. 7.** (a) RF spectrum of the detector output, when biased with -0.8 V and illuminated with mid-infrared Cr:ZnS frequency comb (impulse response); (b) Electrical BW of the Ø121 µm InAs detector measured for a series of bias voltages. Note: -3 dB BW values are the same as in Fig. 6(c), as they were used to fit junction capacitance, while -20 dB BW values are lower due to RF loss at higher frequencies in the real measurement setup.

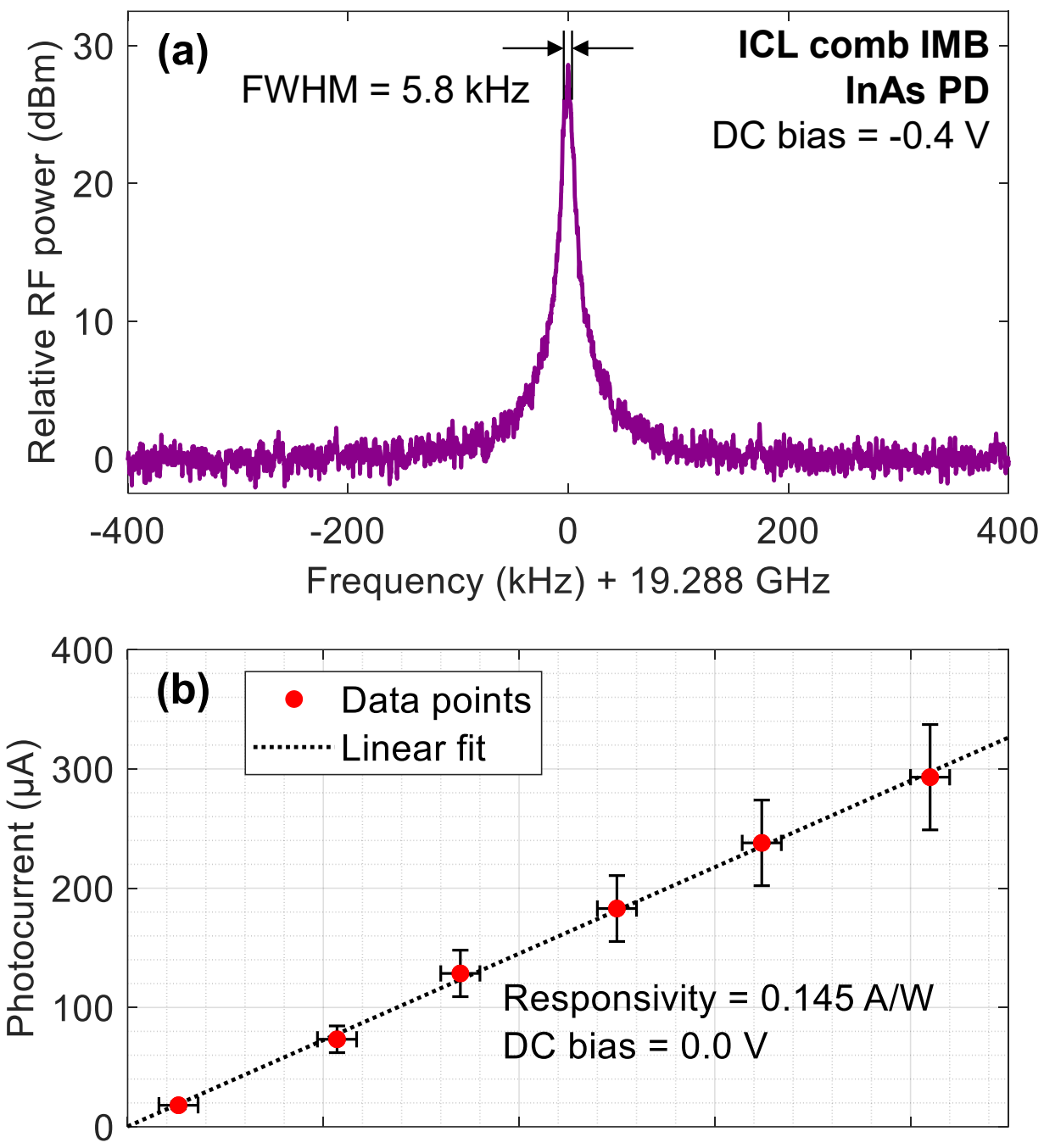

**FIG. 8.** (a) Intermode beat note of the ICL comb operating at 3.25 μm measured with InAs photodetector; (b) measured photocurrent of the detector illuminated with the ICL comb. Note: the incident optical power was measured directly before the focusing lens at the detector input. The error bars indicate the arbitrary precision of the optical power measurement (horizontal) and the accuracy of the ammeter (vertical).

The fast InAs photodetector could also operate far beyond its nominal BW, provided that the detected RF signal was strong enough to make up for pronounced microwave loss. The high responsivity of the InAs detector between 3.20 and 3.45 μm makes it an excellent match for interband cascade lasers (ICLs) operating in this spectral region, known for their applicability in spectroscopy. In particular, the immense speed of the developed InAs photodetector enabled us to measure an intermode beat note (IMB) signal of the ICL frequency comb.[39,40] In Fig. 8(a), we plot an IMB of 2-mm long Fabry-Pérot ICL emitting near 3.25 μm observed at 19.288 GHz with a signal-to-noise ratio of almost 30 dB (at 1 kHz of resolution BW). This demonstration proves the InAs detector operation over an octave further than its −20 dB BW. Figure 8(b), plots the measured photocurrent of the unbiased InAs photodiode when the ICL was used for illumination, yielding a responsivity of 0.145 A/W at 3.25 μm, which agrees well with the data presented in Fig. 2.

In conclusion, we have demonstrated a fast InAs/InAsSb *pBn*-type barrier photodetector operating at GHz speeds. With weak reverse biasing of the photodiode at just −0.8 V, the device achieves an electrical −3 dB BW of 2.4 GHz and −20 dB BW of 8.0 GHz. This is the best optically-confirmed frequency response presented to date in this class of room-temperature MWIR photodetectors. The designed device operates similarly to the *p-i-n* diode by the high voltage drop at the absorber-barrier heterojunction and nearly zero valence band offset between InAs absorber and InAsSbP barrier layer. The remarkable speed of the device was achieved despite not implementing a cascaded structure, T2SL, or UTC mechanism, typically associated with high-speed MWIR photodetectors. It is a notable advantage, taking into account the relative simplicity of an *pBn* barrier detector in terms of fabrication process. The presented performance is also achieved with relatively large Ø121 μm mesas, and can further be improved by downscaling them to diameters typically reported for QCDs and ICIPs (a few tens of microns).[7,10,11] For instance, with a Ø50 μm InAs photodetector mesa, effectively reducing the junction capacitance to 0.3 pF at −0.8 V reverse bias, a −3 dB BW of 11.3 GHz is estimated by our model, although smaller mesas may not follow RC scaling predictions.[41]

Furthermore, the unbiased InAs barrier detector exhibited a responsivity of 0.21 A/W at 3.4 μm. This is a result in majority better than and occasionally comparable to most unbiased room-temperature ICIPs,[9–12,42,43] and UTC detectors presented so far.[16–19,44] Naturally, illumination through the substrate limits the strength and width of the spectral response of the InAs barrier detector to the range of 3.0—3.7 μm. It is noteworthy that due to high substrate refractive index, ~30% Fresnel losses at the air-substrate interface are expected unless anti-reflective coatings are deposited. To further improve the responsivity, the substrate can also be thinned down by etching or lapping. Alternatively, top-illuminating the mesa could greatly enhance and extend the spectral response towards SWIR, with the peak responsivity at 2.5 μm, where extended-InGaAs detectors become slow. Therefore, the development of top-illuminated multi-GHz InAs photodetectors offers an unprecedented opportunity to bridge the spectral gap in both sensitive and fast light detection between SWIR and MWIR.

## ACKNOWLEDGEMENTS

G. Gomółka and Ł. A. Sterczewski would like to thank Dr. Igor Vurgaftman and Dr. Jerry R. Meyer at Naval Research Laboratory (NRL), Washington, DC, USA, for providing the interband cascade laser frequency comb used in this study. G. Gomółka and Ł. A. Sterczewski acknowledge funding from the European Union (ERC Starting Grant, TeraERC, 101117433). Views and opinions expressed are, however, those of the authors only and do not necessarily reflect those of the European Union or the European Research Council Executive Agency. Neither the European Union nor the granting authority can be held responsible for them. Ł. A. Sterczewski and P. Martyniuk acknowledge support from the Swiss Contribution to reducing economic and social disparities in the EU and from the state budget through the National Centre for Research and Development in Poland (SPPW/SWIRLS/0019/2024-00). The project "Ultrastable pulsed lasers covering the spectral range from near to far infrared" (FENG.02.02-IP.05-0069/23) is carried out within the First Team programme of the Foundation for Polish Science co-financed by the European Union under the European Funds for Smart Economy 2021-2027 (FENG). This work was supported by the use of the National Laboratory for Photonics and Quantum Technologies (NPLQT) infrastructure, which is financed by the European Funds under the Smart Growth Operational Programme.

## AUTHOR DECLARATIONS

### Conflict of Interest

The authors have no conflicts to disclose.

### Author Contributions

**Grzegorz Gomółka:** Conceptualization; Data curation; Formal analysis; Investigation; Methodology; Software; Validation; Visualization; Writing – original draft. **Krzysztof Kłos:** Data curation, Formal analysis, Investigation, Resources. **Jarosław Pawluczyk:** Data curation, Formal analysis, Investigation. **Maciej Kowalczyk:** Resources. **Piotr Martyniuk:** Conceptualization, Resources, Supervision, Validation. **Łukasz A. Sterczewski:** Conceptualization; Funding acquisition; Methodology; Project administration; Resources; Supervision; Validation, Writing – review & editing.

## DATA AVAILABILITY

The data that support the findings of this study are openly available in Dataset for "GHz-bandwidth InAs barrier infrared detectors for the 3.0–3.7 μm spectral region operating at room temperature" at
https://doi.org/10.6084/m9.figshare.32346426,
reference number.[45]